\documentstyle{article}
\begin{document}

\title{The Crisis Confronting Standard Big Bang Nucleosynthesis}

\author{Gary Steigman\\
The Ohio State University}

\maketitle

\section{Introduction}

This manuscript is the written version of the material
I attempted to cover at ``Critical Dialogues in Cosmology".  My position is that although BBN is inevitable in
an expanding universe filled with radiation and baryons, at present 
there is a conflict
between the abundances of the light nuclides predicted by standard, big 
bang nucleosynthesis (SBBN) and those inferred from current observational
data.  This challenge to SBBN provides opportunities for astronomy/astrophysics,
cosmology and particle physics.  Perhaps there are unaccounted-for errors in
passing from the observational data to the derived abundances 
(astronomy/astrophysics).
Perhaps there are unaccounted-for systematic corrections in translating the 
derived
abundances from ``here and now" to ``there and then" (astrophysics/cosmology).
Or perhaps the tension between theory and observations is offering us a hint
of new physics beyond the standard model (cosmology/particle physics).

\subsection{The Basics}

Which nuclei may be synthesized during BBN, and in what relative abundances, 
depends
on the interplay among the nucleon density, the radiation temperature and
the early expansion rate.  For example, if the nucleon density were very
small and/or the universe were expanding very rapidly when thermal energies
were comparable to nuclear binding energies, few complex nuclei would have
emerged from the big bang.  In SBBN (isotropy, homogeneity, three flavors of
light neutrinos, etc.) the yields depend on only one ``free" parameter, $\eta$,
the ratio (at present) of nucleons to photons ($\eta = n_{\rm B}/n_{\gamma}$ ; 
$\eta_{10} \equiv 10^{10}\eta$).
In Figure 1 are shown the predicted SBBN yields ($Y_{\rm P}$ is the $^4$He mass
fraction and $y_{2}, y_{3}, y_{7}$ are the ratios by number to H of D,
$^3$He and $^7$Li) as a function of $\eta$.  The present contribution of baryons
to the total density, in units of the critical density, is directly proportional
to $\eta$.  For a present cosmic background radiation temperature of $T_{\rm 
CBR} =
2.73K$,

\begin{equation}  
\Omega_{\rm B} = 0.0146\eta_{10}h_{50}^{-2},
\label{eq:1}
\end{equation}
where the Hubble constant is $H_{0} = 50h_{50}$ km/sec/Mpc.  As may be seen in
Figure 1, the relative abundances vary considerably with $\eta$ so that SBBN
is an overdetermined system in the sense that fixing one of the primordial
abundances fixes $\eta$ and leads to predictions of the abundances of the
three other nuclides.  SBBN is an eminently testable theory!  Consistency
demands that the value of $\eta$ determined by, for example, the primordial
abundance of D lead to predictions for the primordial abundances of $^3$He,
 $^4$He and $^7$Li consistent with the observational data.

\subsection{Status Quo Ante}

Until recently, the program of confronting the predictions of SBBN with
the observational data had led to increasing confidence in the consistency
of the standard model (e.g., WSSOK \cite{WSSOK}).  For example, solar system 
and interstellar (ISM)
observations of D and $^3$He have been used in concert with Galactic evolution
models \cite{ST92,Dearborn-Steigman-Tosi,Palla-etal,Tosi}
(or, nearly model-independent ``inventories" \cite{WSSOK,YTSSO,ST95,D-paper}) to bound the primordial D
abundance from above, leading to a lower bound to $\eta$ ($\eta_{10} \geq 2.8$)
\cite{WSSOK}.
At the same time, observations of $^4$He in low-metallicity, extragalactic HII
regions \cite{Pagel-etal,Skillman-Kennicutt,Skillman-etal,Olive-Steigman-He4}
and of $^7$Li in very metal-poor halo stars 
\cite{Spite-Spite,Thorburn,Molaro-Primas-Bonifacio}
have led to upper bounds on
their primordial abundances, providing an upper bound to $\eta$ ($\eta_{10} 
\leq 4.0$) \cite{WSSOK}.  This quantitative agreement between theory and 
observations for the abundances of the light nuclides, covering 
some nine orders of magnitude, provides strong support for the consistency of 
SBBN.  Furthermore, it should be emphasized that the inferred value (range) 
of $\eta$ is in reasonable agreement with estimates of the baryon
density derived from the dynamics of ``luminous" (i.e., baryonic) matter (see, 
e.g., \cite{SF} and references therein) 
thus providing support for the extrapolation from the present universe to its
earliest epochs.  However, lest we fly too high on the sweet smell of success, 
it is sobering to recount the suggestions of problems with this quantitative 
comparison \cite{crisis-paper}.

\subsection{Hints Of A Crisis}

For SBBN the primordial abundances of D and $^3$He decrease and that of 
$^4$He increases with increasing nucleon-to-photon ratio (see Fig. 1), 
so that the lower 
bound to $\eta$ inferred from observational upper bounds to D and $^3$He leads 
to a predicted lower bound 
to the primordial $^4$He mass fraction.  Alternatively, the upper bound on 
$Y_{\rm P}$ from the observational data leads to an upper bound
to $\eta$ and a lower bound to the primordial abundance of D (or D+$^3$He).
Observations of D and $^3$He ``prefer" a relatively large lower bound to $\eta$
while those of $^4$He ``favor" a relatively small upper bound
to $\eta$.  This ``tension" between D and $^3$He on the one hand, and $^4$He
on the other, provides a hint of a problem 
(e.g., \cite{WSSOK,YTSSO,crisis-paper}).
In Figure 2 this ``crisis" is displayed.  Superposed on the SBBN predictions
for the abundances of D, $^4$He and $^7$Li are
the 68\% and 95\% confidence level bands inferred from the observational data
\cite{crisis-paper}.  Notice
that the range of $\eta$ delimited by D has (at 95\%) no overlap with that
inferred from $^4$He.  This is the ``crisis" \cite{crisis-paper}.  
Another way to visualize the 
crisis is provided by Figure 3 where the 68\% and 95\% CL ranges for $\eta$ 
have been
fixed by the inferred primordial abundances of D, $^3$He and $^7$Li, leading
to predicted ranges for $Y_{\rm P}$ which are to be compared to the upper bound
derived from the HII region data \cite{Olive-Steigman-He4}.  Impressive as is 
the approximate consistency
of SBBN, the standard model is seriously challenged.

\section {Three Possible Solutions}

The hint of a crisis outlined above bears some similarities to several crises
which challenged the 19th century ``standard model" (Newtonian 
mechanics/gravity) when solar system observations appeared to disagree with
theoretical predictions.  The ``Uranus crisis" was real (accurate data) and 
was resolved in favor of the standard model leading to the discovery of 
something 
new: Neptune.  The ``Neptune crisis" was illusory (inaccurate data), the 
standard model remained consistent, and the discovery of Pluto was an 
``accident".  Most exciting, of course, was the resolution of the ``Mercury
crisis".  The crisis was real (the data accurate) and the standard model was
superseded by General Relativity.  Similar options appear available for the
resolution of the current crisis.  Perhaps the data used to derive the 
abundances of one or more of the light nuclides are tainted by unrecognized
statistical or systematic errors (the Neptune crisis).  Or perhaps the data 
are accurate but there are unforeseen astrophysical effects, unaccounted for 
in our extrapolation from the derived abundances to their inferred primordial 
values (the Uranus crisis).  We would, however, be remiss to ignore the
possibility (the Mercury crisis), that this conflict may be revealing 
evidence for new physics (cosmology and/or particle physics).  Although the
ultimate resolution of the current crisis may require some combination of the
above suggestions, here I will consider them separately and identify possible
examples of each.

\subsection {Inaccurate Data: $^4$He ?}

Once the number of light neutrino flavors is fixed ($N_{\nu} = 3$ for SBBN),
the predicted $^4$He mass fraction is a relatively insensitive function of
$\eta$ (see, e.g., Figs. 1-3).  However, as $^4$He is the second most 
abundant nuclide, 
$Y$ may be determined very accurately in a variety of environments spread 
throughout the universe.  One source of the current crisis is the relatively
low upper bound to $\eta$ inferred from the primordial abundance ($Y_{\rm P}$ =
$0.232 \pm 0.003 \pm 0.005$) derived from observations of recombination emission
from hydrogen and helium in extragalactic HII regions 
\cite{Pagel-etal,Skillman-Kennicutt,Skillman-etal,Olive-Steigman-He4}.  
Although there are large numbers of low metallicity HII regions which have 
been observed carefully, minimizing the extrapolation to the primordial
abundance and leading to a
very small statistical uncertainty in $Y_{\rm P}$ (0.003), perhaps there are 
large
systematic corrections along the path from the observed equivalent widths to 
the derived abundances
which exceed the estimate of 0.005 \cite{Olive-Steigman-He4}.  For example, 
since neutral helium is unobservable,
perhaps there is hidden neutral helium in regions where hydrogen is fully
ionized.  If unaccounted for, this would bias the results to abundances which
are systematically low.  The observers (clever as they are) have anticipated
this possibility and have restricted attention to ``high excitation" HII regions
where observations of other ions and comparisons with models suggest this
ionization correction is negligible.  Indeed, for the low metallicity HII
regions observed, the hard radiation spectrum from the very hot stars may even
reverse this effect (neutral H where He is fully ionized) \cite{Skillman-etal}.  
Such an effect
might even correlate with metallicity (downward correction to $Y$ for low
metallicity, upward correction for higher values), leading to a reduction in
the derived value of $Y_{\rm P}$.  Collisional excitation (especially from the
metastable level in HeI) could enhance the helium emission leading to an
overestimate of $Y$.  Most estimates suggest this effect is small, and the
observers do, in one of several ways, try to correct for it.  Perhaps they
have overcorrected, leading to an abundance which is systematically too low.
Again, since collisional excitation is temperature-dependent and metal-poor
HII regions might be hotter, this systematic effect might correlate with
metallicity.

The bottom line here is that if $Y_{\rm P}$ = 0.246 (rather than 0.232;
see, e.g., \cite{Iz-etal}), the 
crisis would be resolved (in favor of a relatively higher value of $\eta$ 
and relatively lower values of primordial D and $^3$He and a higher value
for primordial $^7$Li).

\subsection {Uncertain Extrapolation: D ?}

Until recently, deuterium had only been observed ``here and now" in the solar
system \cite{Geiss} and the (very local) ISM \cite{McCullough,Linsky-etal}
. The older Copernicus UV data and the newer
HST data lead to a reasonably accurate value for the abundance of ISM D
(D/H = $1.6 \pm 0.2 \times 10^{-5}$).  Solar system data (meteoritic $^3$He
abundances, deuterated molecules in the giant planets, etc.) lead to a slightly
larger (and slightly less accurate) abundance \cite{Geiss}.  So far, so good.  
Although
these data are local (in space and in time), they are still of great value 
since the evolution of deuterium is simple.  Other than the big bang, there
are no astrophysical sites for the production of significant abundances of D
\cite{Epstein-Lattimer-Schramm},
and D is destroyed whenever gas forms stars.  Thus, $y_2$ (actually $X_2$,
the deuterium mass fraction) has only decreased
since primordial nucleosynthesis ended ($y_{2\rm P} \geq y_{2\rm ISM}$ ; 
actually, $X_{2\rm P} \geq X_{2\rm ISM}$).  As a result,
observations of deuterium anywhere, anytime provide a lower bound to its
primordial abundance and, therefore, an upper bound to $\eta$.  

The bad news is that both the present ISM and the presolar nebula
are evolved; they contain gas which has been processed through stars (where
D is destroyed).  To bound $\eta$ from below requires that we bound $y_{2\rm P}$
($X_{2\rm P}$) from above and this requires knowledge of Galactic chemical 
evolution.  For
a wide assortment of independent (but similar) chemical evolution models
\cite{ST92,Dearborn-Steigman-Tosi,Tosi,Fields},
designed to account for the observed age-metallicity relation, the observed
gas/stars ratio, various abundance ratios (e.g., secondary to primary nuclei),
etc., it is predicted that D is destroyed by a factor of 2 -- 3 \cite{Tosi}.  
Given the
strong dependence of $y_{2\rm P}$ on $\eta$ (see, e.g., Figs. 1 \& 2), such a 
modest destruction factor coupled with the accurate ISM abundance leads to
a reasonably narrow range for $\eta$ which, however, corresponds to a SBBN
predicted abundance for $^4$He larger than that inferred from the HII region
data (the ``crisis").  Although reasonable, these upper bounds to primordial D
are model-dependent.  Could ``designer" models for Galactic evolution be found
which, while maintaining consistency with the wealth of observational data,
nonetheless permit much larger D destruction?  Without detailed models
which have actually been confronted with the data, it is difficult to answer
this question.  Nonetheless, there are reasons to believe it will not be easy
to find such models based on several nearly model-independent approaches
which have been explored previously.

\subsubsection {The D$ + ^3$He Inventory}

One such approach to bounding D destruction has been to exploit the fact
that when D is incorporated into stars and burned, it is first burned to $^3$He
\cite{Iben,Rood}.
The more resilient $^3$He burns at a higher temperature than D, and so, 
while all the D is destroyed, some
$^3$He survives stellar processing.  Indeed, in stars of all masses there
are interior zones where hydrogen burning results in the production of
new $^3$He.  Thus, in general, the more gas cycled through stars (and the more
D destroyed), the more $^3$He might be expected
\cite{WSSOK,YTSSO,RST,Dearborn-Schramm-Steigman}.  
Unfortunately, the evolution of $^3$He is very complex involving a competition 
between new production and the 
destruction and survival of the prestellar D$+^3$He.  New production of $^3$He 
is uncertain.  If it is ignored current 
observations (ISM and/or solar system) of D and $^3$He \cite{Geiss,McCullough,
Linsky-etal} may be used to provide upper bounds on primordial D and $^3$He, 
nearly independent of the details of Galactic chemical evolution \cite{WSSOK,
YTSSO,ST95,D-paper}.  All the unknown model-dependence of stellar and Galactic
chemical evolution is contained in $\langle g_{3} \rangle$, the average $^3$He 
``survival fraction".
Such an approach was pioneered by Yang et al. \cite{YTSSO} and has been refined 
and updated by Steigman \& Tosi \cite{ST95} and Hata et al. \cite{D-paper}.  For
$\langle g_{3} \rangle$ $\geq 1/4$ and employing solar system D and $^3$He and 
ISM D data, Hata et al. 
\cite{D-paper} derive for SBBN the best fit values ($95\%$ CL): (D/H)$_{\rm P} =
3.5^{+2.7}_{-1.8} \times 10^{-5}$ and $\eta_{10} = 5.0^{+2.9}_{-1.5}$.  This 
$95\%$ CL range for $\eta$, derived from D and $^3$He alone, is considerably 
higher than the previously
preferred range (WSSOK), exacerbating the tension between D and $^4$He since
it predicts (for SBBN) $Y_{\rm P} = 0.247 \pm 0.004$, far in excess of the 
value inferred from the HII region observations \cite{Olive-Steigman-He4}.  
This strong disagreement could 
be ameliorated if $\langle g_{3} \rangle$ is smaller than the value (1/4)
usually adopted \cite{WSSOK,YTSSO,ST95,D-paper,Dearborn-Schramm-Steigman}.  
For $\langle g_{3} \rangle$ $\leq 0.1$, considerably less $^3$He survives, 
permitting more destruction of D, consistent with a larger primordial abundance 
(corresponding to lower values of $\eta$ and $Y_{\rm P}$).  Production of new 
$^3$He, if any, would have an effect similar to that of increasing 
$\langle g_{3} \rangle$, restricting
primordial D and $^3$He to lower abundances leading to higher values of $\eta$.
There are reasons to suspect that current stellar models may overestimate the
production of $^3$He 
 \cite{Dearborn-Steigman-Tosi,Charbonnel,Hogan,Wasserburg-etal,Galli-etal}. 
But, even if total 
destruction in stars less massive than $2.5 M_{\odot}$ is assumed, 
$\langle g_{3} \rangle$ $\geq 0.3$ for gas which has been cycled
through one (and only one) generation of stars \cite{Dearborn-Steigman-Tosi}.  
Thus to reduce $\langle g_{3} \rangle$ 
 would seem to require a model where gas was efficiently cycled through several
generations of stars.  As outlined in the next section, there may be problems 
with such models.

\subsubsection {Constraints On D Depletion From Metallicity And Gas/Stars}

As mentioned above, to relax the upper bound on the primordial abundance of D
(and, correspondingly, the lower bounds on $\eta$ and $Y_{\rm P}$) seems to 
require chemical evolution models which are efficient in cycling gas through
stars.  $^3$He provides one possible constraint on such models which,
however, may be
plagued by uncertainties in stellar modelling.  The observed metallicity
provides another such constraint \cite{Edmunds,ST96}.  The more gas cycled 
through stars, the
higher the metallicity in young stars and newly returned gas.  It will be a
challenge for ``designer" models of Galactic evolution to accomodate large D 
destruction while avoiding overproduction of the heavy elements.  Perhaps
this might be accomplished by expelling (via winds or superbubbles?) the
metal-enriched gas while retaining the D-depleted gas; it remains to be seen
whether realistic models of this type can be found.

Another challenge to such models is the observed ratio of mass in gas to that
in stars (e.g., \cite{Gould-etal}).  Is the relatively high observed gas/stars 
ratio
consistent with models which efficiently cycle the gas through stars (e.g., see
 \cite{Edmunds})?  Such
models might require infall to replenish the gas supply, but if the infalling
gas consists largely of unprocessed material, the ISM D abundance is driven
back towards its primordial value.

Nonetheless, given our ignorance of Galactic chemical evolution, we cannot
dismiss the possibility that D has been destroyed by a large factor 
($\sim$ 5 -- 10) between the
big bang and the present epoch.  If so, lower values of $\eta$ might be
compatible with ISM and solar system observations of D (and $^3$He), leading
to lower predicted values for $Y_{\rm P}$, consistent with the HII region data.
Perhaps the astrophysics is at fault.  Recent data from high redshift (z),
low metallicity (Z) QSO absorbing clouds present ambiguous support for this
possibility (see Sec. 3).

\subsection {New Physics: A Massive Tau Neutrino ?}

Due to the gap at mass-5 and the strong binding of $^4$He, once BBN begins 
most available neutrons are burned to helium-4.  As a result, although $Y_{\rm 
P}$
is relatively insensitive to $\eta$, it is closely tied to the neutron-to-proton
ratio at BBN.  In turn, $n_{\rm n}/n_{\rm p}$ is regulated by the competition 
between
the weak interaction rates ($\rm n \leftrightarrow \rm p$) and the universal 
expansion
rate ($H$).  $H$ is fixed by the total energy density at BBN
and, for SBBN, is dominated by the contribution from the extremely relativistic
particles present ($\gamma, e^{\pm}, \nu_e, \nu_{\mu}, \nu_{\tau}$).  In
units of the photon density, for SBBN, $\rho^{\rm SBBN}_{\rm TOT}/\rho_{\gamma}$
 = 43/8.
If, due to ``new physics", $H$ is changed from its SBBN value, $n_{\rm n}/n_
{\rm p}$
at BBN will be modified leading to a different $Y_{\rm P}$ versus $\eta$ 
relation.  
One such
possibility is that the standard cosmology may be modified due to a variation
in $G$, the Newtonian gravitational ``constant".  Another is that the particle
physics content of the early universe could change due to the presence of
additional light neutrinos or light scalars, or if the tau neutrino were very
massive ($>$ $\sim 10$ MeV) and unstable \cite{Kawasaki-etal}.  It is 
convenient to parameterize such changes by N$_{\nu}$ \cite{SSG} , the 
``effective number of equivalent light neutrinos" (complementary to the number
of standard model neutrinos probed by collider experiments) where N$_{\nu}$ is
defined by,

\begin{equation}  
\rho_{\rm TOT}/\rho^{\rm SBBN}_{\rm TOT} \equiv 1 + 7(N_{\nu} - 3)/43
\label{eq:2}
\end{equation}
For N$_{\nu} > 3$, the early universe expands more quickly, leaving behind
more neutrons to be incorporated into $^4$He during BBN, and vice-versa.  To
a good first approximation, $\Delta{Y_{\rm P}} \approx 0.01(N_{\nu} - 3)$.  
Since
one aspect of the crisis is that SBBN consistent with the inferred primordial
abundances of D, $^3$He and $^7$Li predicts a value for $Y_{\rm P}$ which is 
larger
than that derived from the extragalactic HII regions by  $\Delta{Y_{\rm P}} 
\approx 0.01$, $N_{\nu} \approx 2$ could reestablish consistency
\cite{crisis-paper,2D-paper}.  Of the myriad
possibilities, a massive (unstable) tau neutrino provides one option for
reducing $N_{\nu}$ from its SBBN value of 3 to one closer to 2 
\cite{Kawasaki-etal}.  Current
collider experiments are capable of exploring the interesting mass range
($>$ $\sim 10$ MeV) and either ruling out this option or confirming it.

\section {High-z, Low-Z Deuterium} 

As noted earlier, the relatively strong dependence of the SBBN predicted
abundance of D on $\eta$, coupled with the simple evolution of deuterium,
identifies D/H as the ideal baryometer.  It has been anticipated that
observations of deuterium in high redshift (nearly primordial), low 
metallicity (nearly unevolved) QSO absorption systems would resolve the
current confusion, relieving the tension between theory and observations.
Unfortunately, the few such cases identified to date have added to the
confusion, and rather than relieving the tension, have led to a rupture.
As described eloquently by Hogan \cite{Hogan96}, 
he favors the ``high-D" results ($y_{2\rm P} =
19 \pm 4 \times 10^{-5}$) \cite{Songaila-etal,Rugers-Hogan}
 while I'm sure Tytler \cite{TFB,Burles-Tytler} would argue for the 
``low-D" values ($y_{2\rm P} = 2.4 \pm 0.3 \pm 0.3 \times 10^{-5}$).  As may be 
seen from Figure 4 \cite{2D-paper}, if either 
of these values is correct (rather than something between the two) then the
``true" value of $\eta$ is actually outside the old concordance range ($2.8
\leq \eta_{10} \leq 4.0$).  For example, for ``high-D", $1.3 
\leq \eta_{10} \leq 2.7$ (95\%CL), while for ``low-D", $5.1
\leq \eta_{10} \leq 8.2$ (95\%CL) \cite{2D-paper}.  For ``high-D" the SBBN 
crisis dissipates,
since the predicted abundances of $^4$He and $^7$Li ($Y_{\rm P} = 0.234 \pm 
0.002$,
$y_{7\rm P} = 1.5 \pm 0.6 \times 10^{-10}$) are in excellent agreement with their
inferred primordial values \cite{Dar}.  In this case the challenge is in the 
chemical
evolution court since the comparison of the high primordial D abundance with
the low ISM value requires that deuterium should have been destroyed by a 
factor of $\sim$ 13.  In contrast, for ``low-D", the SBBN crisis is
exacerbated \cite{2D-paper}
($Y_{\rm P} = 0.249 \pm 0.001$, $y_{7\rm P} = 4.7 \pm 0.7 \times 10^{-10}$),
while the D evolution (destruction by a factor of $\sim 1.6$) is consistent
with ``normal" Galactic chemical evolution models (e.g., 
\cite{ST92,Dearborn-Steigman-Tosi,D-paper}.  
As always (and as it should
be) we await more data to resolve the current conundrum.

\section {A Different Path To The Baryon Density}

Given the current ambiguous state of affairs it may be of some value to explore
non-BBN pathways to the baryon density.  Here x-ray clusters may play a valuable
role.  To the extent that rich clusters of galaxies provide a ``fair sample" of
the universal baryon fraction, the cluster baryon fraction may be used in concert with dynamical determinations of the total density (in clustered matter)
to infer the baryon density.

\begin{equation}  
\Omega_{\rm B} = f_{\rm B}\Omega \ \ ; \ \ \eta_{10} = 273f_{\rm B}\Omega h^2
\label{eq:3}
\end{equation}
In equation (3), $H_0 = 100h$ = 70 $\pm$ 10 \cite{Freedman} and $\Omega$ is the
density of clustered matter (= $\Omega_{\rm B} + \Omega_{\rm CDM}$ for ``open"
and ``lambda" CDM models) in units of the critical density.  Within the context
of ``open" and ``lambda" CDM models observations of large scale clustering 
constrain the combination $\Omega h \approx \Gamma \approx 0.25 \pm 0.05$
\cite{Peacock-Dodds}.  From x-ray, optical and ``mini-lensing" studies of rich
clusters, $f_{\rm B}^{\rm Cl} \approx f_{\rm HG} + f_{\rm GAL} \approx 
(0.07 \pm 0.01)h^{-3/2} + (0.02 \pm 0.01)$ so that \cite{sfh} $\eta_{10} 
\approx 6.7 \pm 1.6$.  Actually, further bounds on $\Omega$ and $H_0$ from 
large scale velocity
flows \cite{Dekel} and the age of the Universe \cite{Bolte} tend to increase
$\Omega$ and reduce $H_0$ from the rough estimate presented here, leading to a 
somewhat higher range for $\eta$ \cite{sfh}.  This is  true also for ``mixed"
hot plus cold dark matter models where $\Omega \geq 1 - \Omega_{HDM} \geq 0.7$;
even for $H_{0}$ = 55 $\pm$ 10 \cite{Tammann}, $\eta_{10} \geq 11$ $\pm$ 2.  
Thus, in the context of a variety of CDM models,
the preliminary results of this ``dynamical" approach to the baryon density
suggest a high value for $\eta$, consistent with that inferred from solar
system and ISM D and $^3$He \cite{D-paper}, favoring the low-D/high-$^4$He
option.

\section{Acknowledgments}

I owe a debt of gratitude to my brothers in BBN (S. Bludman, N. Hata, 
P. Langacker, R. J. Scherrer, D. Thomas and T. P. Walker) from whom I've
learned so much.  It is mainly our joint work which is described here.  I
also wish to acknowledge D. Dearborn and M. Tosi for advice and guidance
on stellar and Galactic evolution. I am especially grateful to J. E. Felten
for his careful reading of this manuscript and his many valuable suggestions
and corrections.

\vfill\eject

\centerline{\bf Figure Captions}

Figure 1.   SBBN predicted primordial abundances of $^4$He 
(mass fraction $Y_{\rm P}$), D ($y_{2\rm P}$ = (D/H)$_{\rm P}$), $^3$He 
($y_{3\rm P}$ = 
($^3$He/H)$_{\rm P}$) and $^7$Li ($y_{7\rm P}$ = ($^7$Li/H)$_{\rm P}$) as a 
function of
$\eta$, the nucleon-to-photon ratio.  This graph has been provided by D. Thomas.

Figure 2.   SBBN predictions (solid lines) for $Y_{\rm P}$, $y_{2\rm P}$ and 
$y_{7\rm P}$
along with their theoretical uncertainties ($1\sigma$) estimated via Monte
Carlos (dashed lines) \cite{crisis-paper}.  Also shown are the regions 
constrained by the observations at 68\% and 95\% (shaded regions and dotted
lines respectively).

Figure 3.   Theoretical (SBBN) predictions for the primordial helium-4 mass 
fraction ($Y_{\rm th}$) at 68\%
(shaded region) and 95\% (dotted curve) for $\eta$ constrained by the inferred
primordial abundances of D, $^3$He and $^7$Li.  Also shown are the 68\% and
95\% bounds to $Y_{\rm P}$ inferred from observations of low-metallicity,
extragalactic HII regions ($Y_{\rm obs}$).  The absence of overlap between 
Y$_{\rm obs}$ and Y$_{\rm th}$ is the ``crisis".

Figure 4.   As for Figure 2 but, with the deuterium abundance inferred from
the two sets of QSO absorption data \cite{2D-paper}.


\begin{thebibliography}{}   

\bibitem{WSSOK}
Walker, T. P., Steigman, G., Schramm, D. N., Olive, K. A., \& Kang, H. 1991,
ApJ, 376, 51

\bibitem{ST92}
Steigman, G.  \& Tosi, M. 1992, ApJ, 401, 15 

\bibitem{Dearborn-Steigman-Tosi}
Dearborn, D., Steigman, G. \& Tosi, M. 1996,
ApJ, 465, 887

\bibitem{Palla-etal}
Palla, F., Galli, D. \& Silk, J. 1995,
ApJ, 451, 44

\bibitem{Tosi}
Tosi, M., 
in {\it From Stars to Galaxies},
edited by C.\ Leitherer, U.\ Fritze von Alvensleben, and J.\ Huchra
(ASP Conference series, 1996).

\bibitem{YTSSO}
Yang, J., Turner, M. S., Steigman, G., Schramm, D. N. \& Olive, K. A. 1984,
ApJ, 281, 493

\bibitem{ST95}
Steigman, G. \& Tosi, M. 1995,
ApJ, 453, 173

\bibitem{D-paper}
Hata, N., Scherrer, R. J., Steigman, G., Thomas, D. \& Walker, T. P. 1996,
ApJ, 458, 637

\bibitem{Pagel-etal}
Pagel, B. E. J., Simpson, E. A., Terlevich, R. J. \& Edmunds, M. G. 1992,
Mon. Not. Roy. Astron. Soc., 225, 325

\bibitem{Skillman-Kennicutt}
Skillman, E. D. \& Kennicutt, R. C. 1993,
ApJ, 411, 655

\bibitem{Skillman-etal}
Skillman, E. D., Terlevich, R. J., Kennicutt, R. C., Garnett, D. R. \&
Terlevich, E. 1994,
ApJ, 431, 172

\bibitem{Olive-Steigman-He4}
Olive K. A. \& Steigman, G. 1995,
ApJS, 97, 49

\bibitem{Spite-Spite}
Spite, F. \& Spite, M. 1982,
Astron. Astrophys., 115, 357

\bibitem{Thorburn}
Thorburn, J. A. 1994,
ApJ, 421, 318

\bibitem{Molaro-Primas-Bonifacio}
Molaro, P., Primas, F. \& Bonifacio, P. 1995,
Astron. Astrophys., 295, 47

\bibitem{SF}
Steigman, G. \& Felten, J. E. 1995,
Space Science Reviews, 74, 245

\bibitem{crisis-paper}
Hata, N., Scherrer, R. J., Steigman, G., Thomas, D., Walker, T. P., 
Bludman, S. \& Langacker, P. 1995,
Phys. Rev. Lett., 75, 3977

\bibitem{Iz-etal}
Izotov, Y. I., Thuan, T. X. \& Lipovetsky, V. A. 1994,
ApJ, 435, 647

\bibitem{Geiss}
Geiss, J. 1993,
in {\it Origin and Evolution of the Elements},
edited by N.\ Prantzos, E.\ Vangioni-Flam, and M.\ Casse
(Cambridge University Press, Cambridge, 1993), p.\ 89.

\bibitem{McCullough}
McCullough, P. R. 1992, 
ApJ, 390, 213

\bibitem{Linsky-etal}
Linsky, J. L. {\it et al}. 1993,
ApJ, 402, 694

\bibitem{Epstein-Lattimer-Schramm}
Epstein, R., Lattimer, J., \& Schramm, D. N. 1976,
Nature, 263, 198

\bibitem{Fields}
Fields, B. D. 1996,
ApJ, 456, 478

\bibitem{Iben}
Iben, I. Jr. 1967,
ApJ, 147, 624

\bibitem{Rood}
Rood, R. T. 1972,
ApJ, 177, 681

\bibitem{RST}
Rood, R. T., Steigman, G. \& Tinsley, B. M. 1976,
ApJ, 207, L57

\bibitem{Dearborn-Schramm-Steigman}
Dearborn, D. S. P., Schramm, D. N. \& Steigman, G. 1986,
ApJ, 302, 35

\bibitem{Charbonnel}
Charbonnel, C. 1995,
ApJ, 453, L41

\bibitem{Hogan}
Hogan, C. J. 1995,
ApJ, 441, L17

\bibitem{Wasserburg-etal}
Wasserburg, G., Boothroyd, A. \& Sackmann, I.-J. 1995,
ApJ, 447, L37

\bibitem{Galli-etal}
Galli, D., Palla, F., Ferrini, F. \& Penco, U. 1995,
ApJ, 443, 536

\bibitem{Edmunds}
Edmunds, M. G. 1994,
Mon. Not. Roy. Astron. Soc., 270, L37

\bibitem{ST96}
Steigman, G. \& Tosi, M. 1996,
in preparation

\bibitem{Gould-etal}
Gould, A., Bahcall, J. N. \& Flynn, C. 1996,
ApJ, 465, 759

\bibitem{Kawasaki-etal}
Kawasaki., M., Kernan, P., Kang, H.-S., Scherrer, R. J., Steigman, G. \& Walker,
T. P. 1994,
Nuc. Phys. B, 419, 105

\bibitem{SSG}
Steigman, G., Schramm, D. N. \& Gunn, J. E. 1977,
Phys. Lett. B, 66, 202

\bibitem{Hogan96}
Hogan, C. J. 1996,
This Volume

\bibitem{Songaila-etal}
Songaila, A., Cowie, L. L., Hogan, C. J. \& Rugers, M. 1994,
Nature, 368, 599

\bibitem{Rugers-Hogan}
Rugers, M. \& Hogan, C. J. 1996,
ApJ, 459, L1

\bibitem{TFB}
Tytler, D., Fan, X. M. \&  Burles, S. 1996,
Nature, 381, 207

\bibitem{Burles-Tytler}
Burles, S. \& Tytler, D. 1996,
Science, submitted (astro-ph/9603070)

\bibitem{2D-paper}
Hata, N., Steigman, G., Bludman, S. \& Langacker, P. 1996,
Phys. Rev. D, in press (astro-ph/9603087)

\bibitem{Dar}
Dar, A. 1995,
ApJ, 449, 550

\bibitem{Freedman}
Freedman, W. 1996,
This Volume

\bibitem{Peacock-Dodds}
Peacock, J. A. \& Dodds, S. J. 1994,
Mon. Not. Roy. Astron. Soc., 267, 1020

\bibitem{sfh}
Steigman, G., Felten, J. E. \& Hata, N. 1996,
In preparation (OSU-TA-14/96)

\bibitem{Dekel}
Dekel, A. 1994,
Ann. Rev. Astron. Astrophys., 32, 371

\bibitem{Bolte}
Bolte, M. 1996,
This Volume

\bibitem{Tammann}
Tammann, G. A. 1996,
This Volume 

\end{thebibliography}
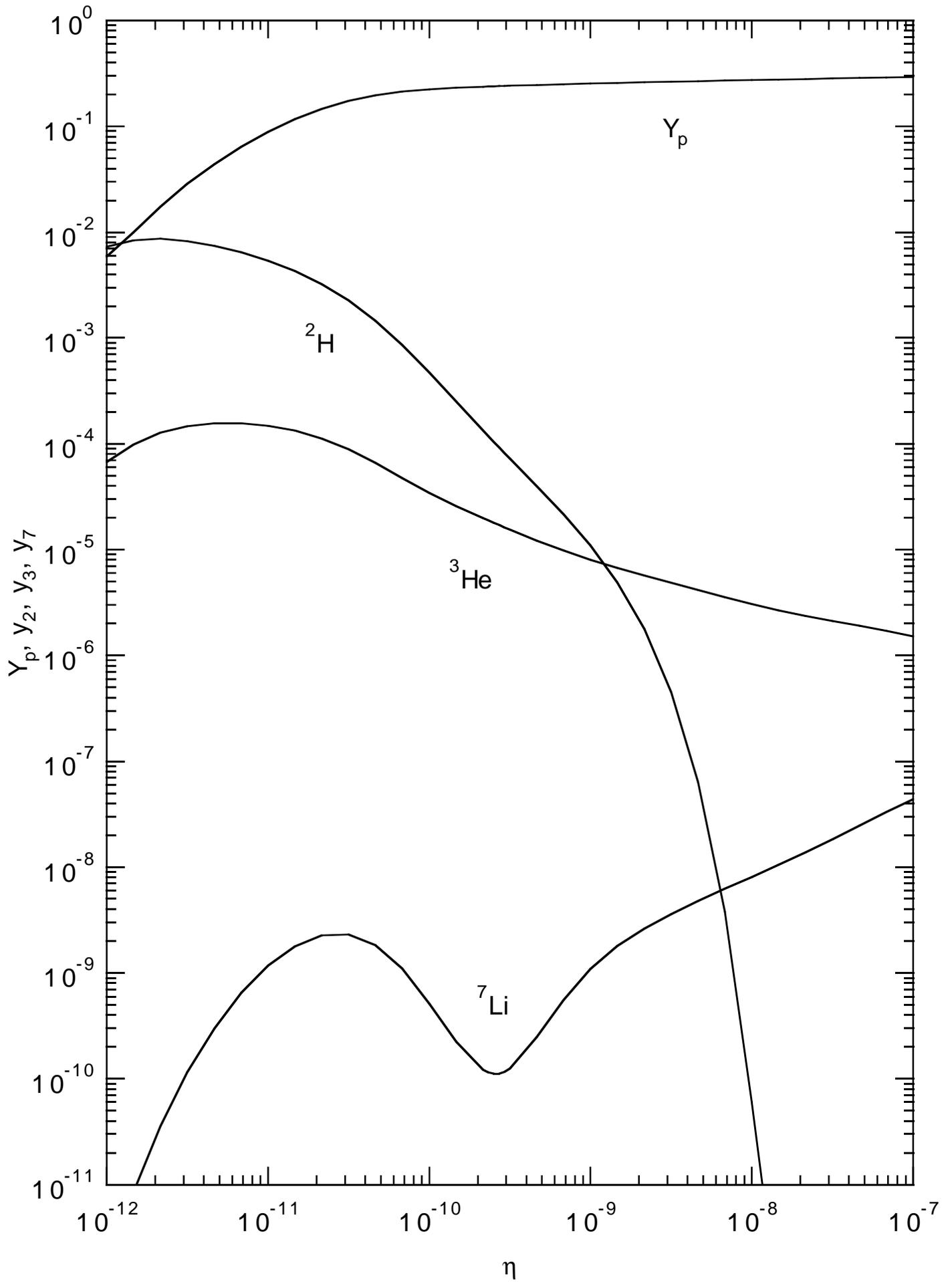
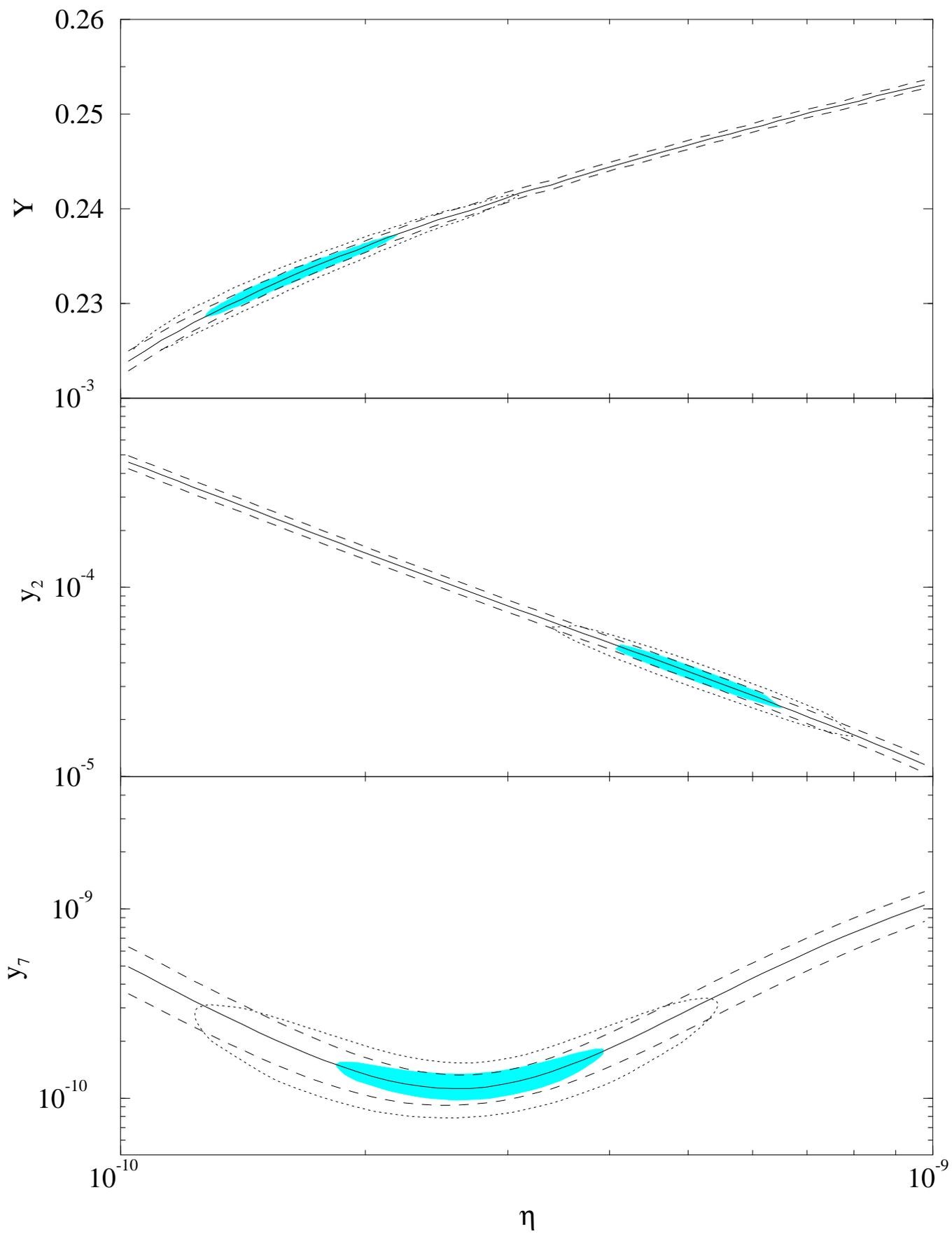
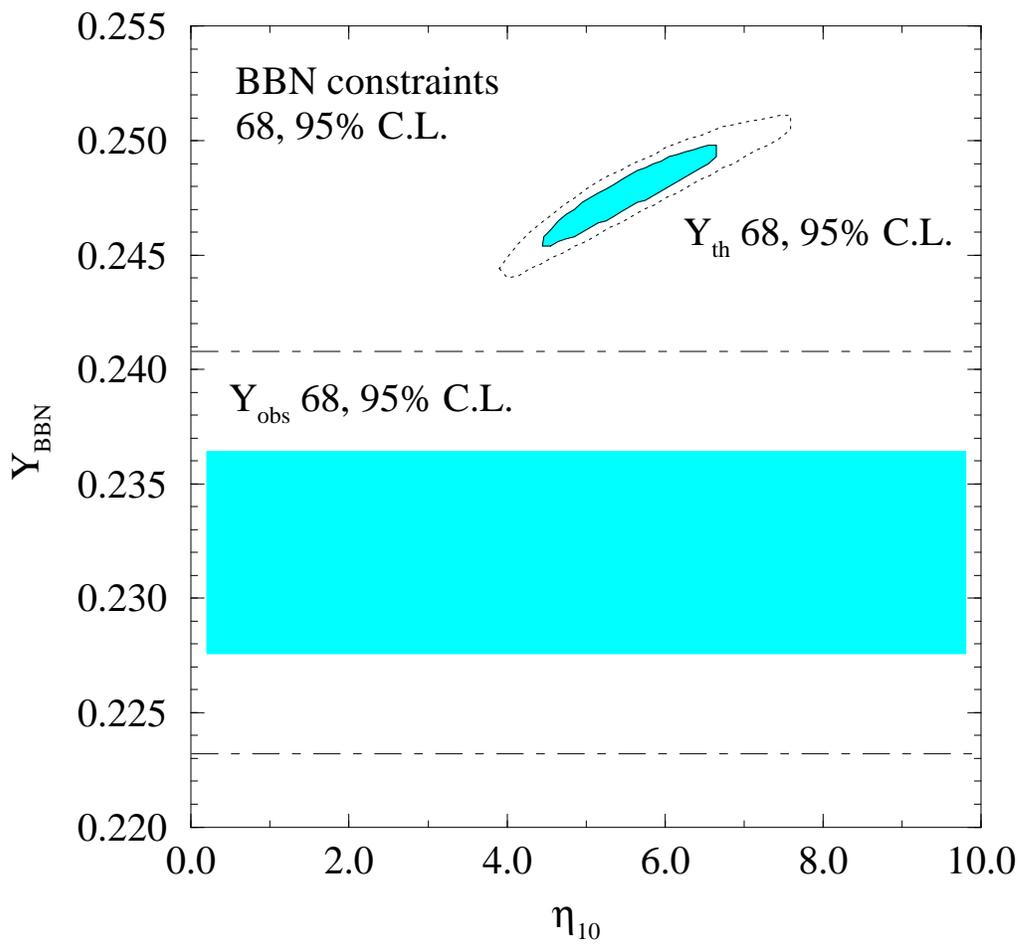
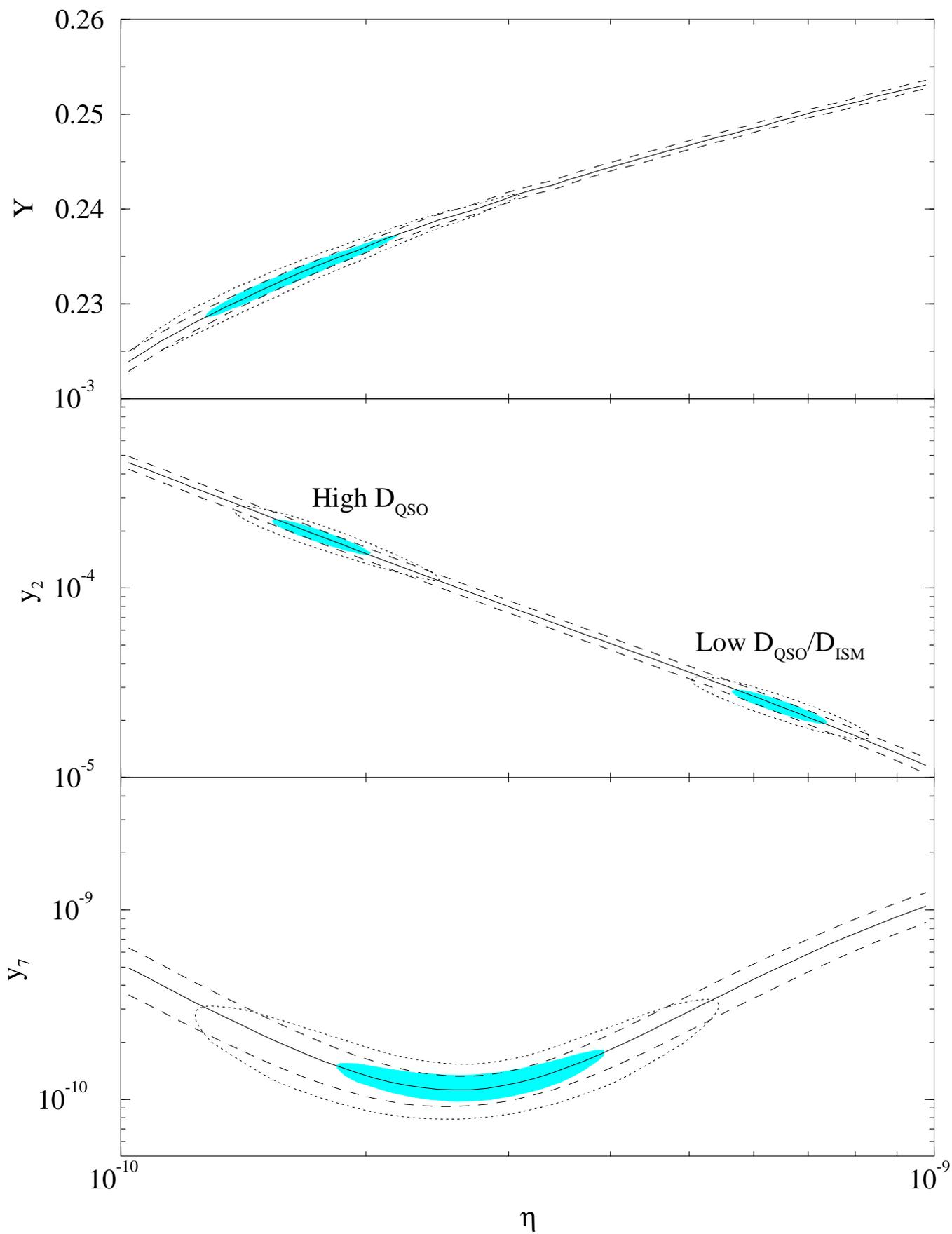
\end{document}